\documentstyle[twocolumn,eqsecnum,aps]{revtex}
\def\be{\begin{equation}}
\def\ee{\end{equation}}
\def\bea{\begin{eqnarray}}
\def\eea{\end{eqnarray}}
\begin{document}
\draft
\title{Dynamical vertex mass generation and chiral symmetry breaking
on the light-front}
\author{Matthias Burkardt}
\address{Department of Physics\\
New Mexico State University\\
Las Cruces, NM 88003-0001\\U.S.A.}
\maketitle
\begin{abstract}
Naively, helicity flip amplitudes for fermions
seem to vanish in the chiral limit of
light-front QCD, which would make it nearly impossible to generate a
small pion mass in this framework. Using a simple model, it is illustrated
how a large helicity flip amplitude is generated dynamically by summing
over an infinite number of Fock space components. 
While the kinetic mass is basically generated by a zero-mode induced
counter-term, the vertex mass is generated dynamically by infinitesimally
small, but nonzero, momenta.
Implications for the
renormalization of light-front Hamiltonians for fermions are discussed.
\end{abstract}
\section{Introduction}
Deep inelastic lepton-hadron scattering has played a
fundamental role in the investigation of hadron
structure. For example, the discovery of Bjorken scaling
confirmed the existence of point-like charged
objects inside the nucleon (quarks). 
Besides such fundamental discoveries, deep inelastic
scattering (DIS) 
revealed surprising and interesting details
about the structure of nucleons and nuclei, such as
the nuclear EMC effect,
the spin crisis in polarized DIS experiments 
and
the isospin asymmetry of the nucleon's Dirac sea.

Perturbative QCD evolution has been successfully applied to
correlate large amounts of experimental data.
However, progress in understanding non-perturbative
features of the parton distributions in these experiments
has been slow. In fact, the theoretical understanding of
the surprising results listed above is mostly limited
to {\it ad hoc} models with little connection to the
underlying quark and gluon degrees of freedom.

Part of the difficulty in describing parton distributions
nonperturbatively derives from the fact that parton
distributions measured in DIS are dominated by
correlations along the light-cone ($x^2=0$).
For example, this makes calculations of parton distributions on
a Euclidean lattice, where all distances are
space-like, very difficult.
Furthermore, in an equal time quantization scheme,
deep inelastic structure functions are described by
real time response functions which are not only
very difficult to interpret but also to calculate.

Light-front (LF) quantization provides the most physical approach towards
calculating the quark-gluon structure of hadrons measured in
deep-inelastic lepton-nucleon scattering experiments 
\cite{osu,bigguy,mb:adv}. Furthermore, LF quantization seems 
to be a promising
tool to describe the immense wealth of experimental
information about structure functions since
correlations along the light-cone become ``static'' observables
in this approach
[i.e. equal $x^+ \equiv (x^0+x^3)/\sqrt{2}$ observables]. 
This implies that parton distribution functions are easy to 
evaluate from the LF wavefunctions and are easily interpreted as LF 
momentum densities. Similar simplifications apply to many other
high-energy scattering observables, including non-forward
parton distributions measurable in deeply virtual Compton
scattering experiments \cite{dvcs}.

Further advantages of the LF formalism derive from the
simplified vacuum structure (nontrivial vacuum effects
can only appear in zero-mode degrees of freedom) which
provides a physical basis for the description of
hadrons that stays close to intuition:
fields are quantized at equal LF-time $x^+=(x^0+x^3)/\sqrt{2}$.
As a result, the longitudinal momentum $p^+$ for all quanta is
both a kinematical operator (no interactions) and strictly
positive (as long as zero-modes, i.e. modes with $p^+$ strictly zero,
are not explicitly included as dynamical degrees of freedom).
Since the Fock-vacuum has vanishing $p^+$ momentum, this implies that
interactions cannot mix states that contain particle hole excitations on 
top of the Fock vacuum
with the Fock vacuum itself, and hence the full vacuum is the Fock vacuum.
The resulting apparent contradiction between nontrivial vacua in a 
conventional formulation of quantum field theory and trivial vacua on the LF 
is resolved upon realizing that one should not merely omit zero-mode
degrees of freedom from LF Hamiltonians but rather integrate them out
which results in the concept of effective LF-Hamiltonians 
\cite{osu,mb:adv,nato}. In such an approach the vacuum structure is shifted from
the states to the operators (e.g. the Hamiltonian) and one can 
(at least in principle) account for nontrivial
vacuum structure in the renormalization procedure.

Constructing effective LF Hamiltonians is in general a nontrivial task. 
For a variety of model field theories, this task has
been successfully accomplished (examples can be found in Refs.
\cite{osu,mb:adv,mb:hala,mb:sg,mb:yuk}) and as far as theories
with fermions is concerned, the following general features emerge:

In the LF-formulation, only half of the spinor components are dynamical
degrees of freedom in the sense that their equation of motion 
involves a time derivative. Upon introducing 
$\psi_{(\pm)}=\gamma^\pm \gamma^\mp \psi/2$, one finds for example in
QCD that $\psi_{(-)}$ satisfies a constraint equation \cite{osu}
\begin{equation}
i\partial_- \psi_{(-)} = \left[ {\vec \alpha}_\perp \cdot
\left( i{\vec \partial}_\perp + g {\vec A}_\perp\right) + \gamma^0m_F
\right] \psi_{(+)}
\end{equation}
and $\psi_{(-)}$ is usually eliminated (using this constraint equation)
from the Lagrangian before quantizing the theory.
Thus the Hamiltonian contains both a term quadratic in the fermion mass
(the kinetic energy term for the fermions) and one term which is linear
in the fermion mass (one gluon vertex with helicity flip).

It has been known for a long time that integrating out zero-mode
degrees of freedom results in a nontrivial renormalization of the
quadratic (kinetic!) mass term \cite{mb:yuk,mb:rot} 
but the linear (helicity flip 
vertex!) mass term in the Hamiltonian is unaffected by strict zero modes
and the ``vertex mass'' must be identified with the current quark mass
which vanishes in the chiral limit \cite{osu,mb:hala,ken}.

It is very easy to see how a constituent quark picture can emerge in such
an approach.
However, it always seemed mysterious how one can obtain a massless $\pi$
meson in such a picture without having at the same time a massless $\rho$:
when the helicity flip term for quarks is omitted, $\pi$ and $\rho$ become
partners in a degenerate multiplet \cite{ken}. The key observation to resolve
this problem is that one needs to find a mechanism which dynamically 
generates a large helicity flip amplitude. At first this seems impossible
since every LF-time ordered diagram (any order in the coupling!) which flips 
the helicity of the fermion contains at least one power of 
the vertex mass (which vanishes in the chiral limit).
In this paper, we will consider a toy model in which we explicitly
demonstrate that summing over all orders in (Hamiltonian) perturbation theory 
yields a large helicity flip amplitude even though every single term
in the perturbation series vanishes in the chiral limit.

\section{A Model with Dynamical Chiral Symmetry Breaking}
The above argument does not rule out the possibility that summing over
an infinite class of time-ordered diagrams (including an infinite
number of Fock space components) can lead to divergences
which can compensate for the suppression of individual diagrams.
A simple  model to illustrate this idea is a fermion field with
(fundamental) ``color'' degrees of freedom coupled to the transverse 
components of a massive vector field (adjoint representation) 
\begin{equation}
{\cal L} = \bar{\psi}\left( i \partial\!\!\!\!\!\!\not \;\; - m -
g\sqrt{\frac{\pi}{N_C}}
{\vec \gamma}_\perp {\vec A}_\perp \right)\psi - \frac{1}{2}
\mbox{tr} \!\!\left( {\vec A}_\perp \Box {\vec A}_\perp + 
\lambda^2 {\vec A}_\perp^2\right).
\label{eq:model}
\end{equation}
We will take the limit $N_C\rightarrow \infty$, where
the planar approximation becomes exact. The factor of 
$\sqrt{\frac{\pi}{N_C}}$ has been included in the coupling for later 
convenience.
It should be emphasized that even though the interactions in Eq.
(\ref{eq:model}) resemble QCD to some extent, we regard
Eq. (\ref{eq:model}) as neither an approximation to QCD nor a phenomenological
model for hadrons. Instead we take Eq. (\ref{eq:model}) as a toy
model with interactions that also appear in QCD, and the motivation
to study this model is to shed some light on the mechanism for
chiral symmetry breaking in the LF framework in general. 

The reason for choosing the model in this particular way
is as follows: using a model with an infinite number of ``colors'' (and
no self-interaction for the boson field) renders the model
solvable since the ladder-rainbow approximation becomes exact.
A vector coupling of the bosons to the fermions is chirally invariant,
and restricting the interactions to the transverse components
was done to facilitate comparisons between ET and LF solutions to
the model.

This model has been introduced and studied in Ref. \cite{mb:hala}, where
it has been shown to exhibit dynamical chiral symmetry breaking
(D$\chi$SB) by studying the Euclidean Dyson-Schwinger (DS) equations
for the model. 
Furthermore, it was shown that, for example, when a
transverse momentum cutoff is used both in the ET as well as the LF
formulation then both formulations are equivalent to all orders in
perturbation theory provided the current mass in the ET framework
is identified with the vertex mass \footnote{This is the mass term
which appears in the helicity flip coupling in the LF Hamiltonian.
Below we will define all couplings in the Hamiltonian in an approximation
to the model.} and if an appropriate counter-term (which arises from
integrating out zero-mode degrees of freedom) is
added to the kinetic mass term \footnote{This is the quadratic mass term which
appears in the fermion kinetic energy term in the LF Hamiltonian.}.
This result clearly shows that while integrating out zero-modes 
causes kinetic mass generation for the fermions, it does not
modify the only spin dependent term in the LF Hamiltonian
(the helicity flip vertex which has a coupling proportional to the 
current mass). Thus fermion spin degrees of freedom seem to decouple in
the chiral limit. If spin would really decouple, then the $\pi$ and
$\rho$ (any helicity) should become degenerate, i.e. either 
both the $\pi$ and the $\rho$ should be massless or both be massive.
Both possibilities would contradict the ET calculation (which has
been shown to be equivalent to the LF calculation to all orders)
where one can prove Goldstone's theorem explicitly.

The resolution to this apparent paradox is that the above argument about 
the supposed degeneracy of $\pi$ and $\rho$ contains a flaw:
while helicity flip amplitudes for {\it bare} fermions are indeed
proportional to the current mass, it is conceivable that ``dressing
the fermions'', i.e. summing over an infinite number of Fock space
components, leads to divergences in the chiral limit, which cancel
the vanishing of the helicity flip coupling and lead to finite helicity flip
for physical fermions and hence a nonzero $\pi$-$\rho$ splitting.

In this paper, we will investigate a particular approximation to the
above model [Eq. (\ref{eq:model})], and we will demonstrate explicitly
that, even with a helicity flip vertex proportional to the current mass,
finite helicity flip amplitudes are obtained for physical
states in the chiral limit and that it is crucial to sum over an
infinite number of Fock components in order to obtain this result.

\section{Dynamical Vertex Mass Generation}

Even though the rainbow approximation is solvable in any number of
dimensions, we will restrict ourselves to a dimensionally reduced
version of the above model (\ref{eq:model}), i.e. we will assume that 
the fields depend on longitudinal coordinates only,
but still have the full four component $\gamma$-matrix structure.

Canonical quantization thus yields for the 
LF-Hamiltonian of the model
\begin{eqnarray}
H &=& \sum_h\int_0^\infty \!\!\!dk
\left[ \frac{m_{kin}^2}{2k} b_{k,h,\alpha}^\dagger b_{k,h,\alpha}
+ \frac{\lambda^2}{2k} a_{k,h,\alpha,\beta}^\dagger a_{k,h,\alpha,\beta} 
\right] \label{eq:h}
\\
& &+ \frac{g}{2\sqrt{N_C}} 
\sum_h \int_0^\infty \!\!\!dp dq \frac{dk}{\sqrt{k}}
b_{p,h,\alpha}^\dagger \left(\frac{m_V}{p}-\frac{m_V}{q}\right)
\times \nonumber\\ & &  
\left[ \delta(p-q-k) a_{k,h,\alpha,\beta} - \delta(p+k-q) 
a^\dagger_{k,-h,\alpha,\beta} \right] b_{q,-h,\beta}
\nonumber\\
& &+ \frac{g^2}{2N_C} 
\sum_h \int_0^\infty \!\!\! dp dq \frac{dk dr}{\sqrt{kr}} 
b_{p,h,\alpha}^\dagger 
\times \nonumber\\ & & \quad \quad \quad 
\left[ \frac{\delta (p+k-q-r)}{p+k} a_{k,-h,\alpha,\beta}^\dagger 
a_{r,-h,\beta,\gamma}\right.
\nonumber\\
& & \quad\quad\quad\quad + 
\frac{\delta(p+k+r-q)}{p+k} a_{k,-h,\alpha,\beta}^\dagger
a^\dagger_{r,h,\beta,\gamma}
\nonumber\\ & & \quad 
\quad \quad \quad +\left.
\frac{\delta(p-k-r-q)}{r+q} a_{k,h,\alpha,\beta} a_{r,-h,\beta,\gamma}\right]
b_{q,h,\gamma}
\nonumber
\end{eqnarray}
where $h$ labels the helicity of the fermion/boson and Greek indices represent
color (note that one can also drop color indices completely
as long as one keeps track of the color ordering in the Fock space). 
Because of the large $N_C$ limit and because we are studying the
vertex of a fermion in this paper, quark pair creation can be
neglected and all terms involving antiquarks are not shown in
Eq. (\ref{eq:h}).

Helicity is conserved at each vertex in this dimensionally reduced 
model, which implies that fermion helicity flips in each one boson emission 
or absorption process and the helicity of the emitted boson will have 
the same sign as the helicity of the fermion before the emission.
In the large $N_C$ limit, bosons dressing a fermion must always be
absorbed in reverse order to the order in which they have been
emitted. Thus it makes sense to label the bosons in a given Fock
state in the order in which they have been produced. This is
the most convenient way to represent a state in this model. In particular, 
using the abovementioned helicity selection rules for this 
model, one can easily convince oneself that for any state which mixes
with a bare fermion, the sign of the helicity for the bosons alternates
as one moves along the state (which is ordered using the rule 
described above).
Thus it is not necessary to keep track of helicity degrees of freedom
explicitly, and we will completely omit helicity labels from now on.
\footnote{There are two equivalent ways how one can recover the hidden
helicity information for a dressed fermion: from the wave function by
counting the number of ``gluons'' (an odd number means the helicity
of the fermion is reversed to its helicity in the bare Fock component),
or by counting the number of ``gluons'' in an intermediate state in 
perturbation theory.}

In the canonical Hamiltonian there is only one fermion mass $m_{kin}=m_V=m$,
but in general, when zero-mode degrees of freedom have been integrated
out, one has to distinguish between $m_{kin}$ and $m_V$. \footnote{The only
exception seem to be calculations where a Pauli-Villars regulator is used
(see Refs. \cite{mb:rot,hiller} and references therein).}

The only interaction term in Eq. (\ref{eq:h}) which flips the helicity
of the quarks is the 3-point coupling, which is proportional to $m_V$.
\footnote{Note that total helicity ($h_q+2h_g$) is conserved at each vertex, 
which follows from the fact that setting the $x$ and $y$ components of 
all momenta equal to zero is rotational invariant about the $z$ axis.}

In the following, we will study corrections to the vertex of the
$A_\perp$-field in Hamiltonian perturbation theory. This analysis
is greatly simplified due to the large $N_C$ limit for the following
two reasons: First of all, the only
``vertex corrections'' are actually given by diagrams that resemble
a self energy correction to the external legs which are connected
to the ``external'' vertex by an instantaneous interaction.
Secondly, the rainbow approximation for self-energies becomes exact.

In order to organize the perturbative expansion, it is useful to introduce
some notation and graphical symbols representing the notation in
time-ordered diagrams.
First we introduce the fermion Green's function $G(p,E)$. For a free theory
$G(p,E)=1/(E-\frac{m^2}{2p}+i\varepsilon)$, while for an interacting theory
\begin{equation}
G(p,E) = \sum_n \frac{p_n(1)}{E-E_n+i\varepsilon} ,
\end{equation}
where the sum is over eigenstates of the Hamiltonian with momentum $p$
and energy eigenvalue $E_n$ and where $p_n(1)$ denotes the probability
to find the n-th state in the lowest Fock component.
In diagrams, all propagating (i.e. non-instantaneous) fermion lines
are implicitly assumed to represent the full Green's function $G(p,E)$.
\begin{figure}
\unitlength1.cm
\begin{picture}(15,7.5)(1,-8)
\includegraphics{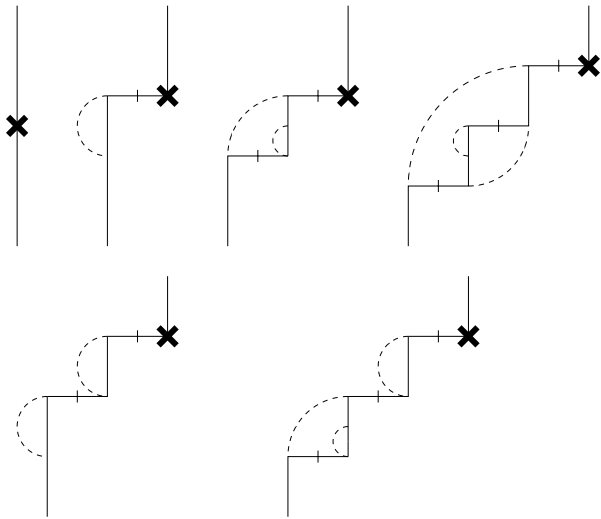}
\end{picture}
\caption{Typical LF-time ordered diagrams contributing to the helicity 
flip amplitude (one ``gluon'' vertex) for a fermion in planar approximation. 
The external gluon vertex is depicted by a cross.
The dashed lines are boson fields and the slashed 
lines represent LF-instantaneous interactions. Since the instantaneous
interactions conserve fermion helicity, the actual helicity flip occurs
at the non-instantaneous vertex inside the loop corrections.
Only diagrams with corrections to the incoming fermion line are drawn,
but the corrections might instead occur on the outgoing line.
}
\label{fig0}
\end{figure}
Vertex corrections in the model can only occur on
either the incoming or the outgoing line, but not on both at the same
time. \footnote{With the exception of theories that contain couplings to
``bad currents'' (such as a coupling involving $\bar{\psi} \gamma^- \psi$)
there cannot be two instantaneous interactions couple to the same
vertex. In the above example, this can be seen explicitly by using
$\gamma^+/p^+$ for the instantaneous propagator and by using the
$\gamma$-algebra $\gamma^+ \gamma_\perp \gamma^+=0$.}
It thus makes sense to decompose the full vertex function 
$\Gamma (p,p^\prime,E)$
(denoted by a full blob) into the bare piece, corrections to
the incoming line $\Gamma_i(p,E)$ (denoted by a full semicircle
on the incoming line and connected via an instantaneous interaction 
to the external vertex) as well as corrections to the outgoing line
$\Gamma_o(p^\prime,E)$ (denoted by a full semicircle
on the outgoing line and connected via an instantaneous interaction 
to the external vertex). \footnote{Actually, 
there is a very simple relation between vertex corrections
on the incoming and outgoing lines,
$
\Gamma_i(p,E)= - \Gamma_o(p,E),
$
but we found it useful to distinguish them in order to better 
elucidate the details of the recursive relations.}
 
\begin{equation}
\Gamma (p,p^\prime,E,E^\prime) = \left(\frac{m_V}{2p} - 
\frac{m_V}{2p^\prime}\right)+\Gamma_o (p^\prime,E^\prime)
+\Gamma_i (p,E),
\label{eq:g1}
\end{equation}
\begin{figure}
\unitlength1.cm
\begin{picture}(15,3)(1.2,-4.5)
\includegraphics{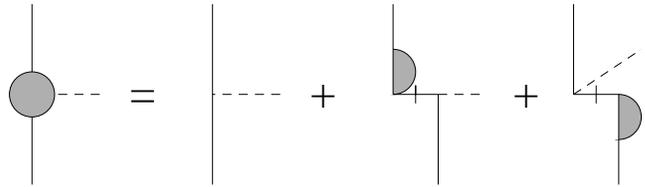}
\end{picture}
\caption{
Decomposition of the full vertex into the bare vertex, corrections 
attached to the outgoing line and corrections
to the incoming line. 
}
\label{fig1}
\end{figure}
where $E$ and $E^\prime$ are related by energy conservation.
They satisfy the integral relations 
\begin{eqnarray}
\Gamma_i(p,E) &=& \frac{g^2}{p} \int_0^p \!\!\! \frac{dk}{2k}
G(p-k,E-\frac{\lambda^2}{2k}) \times \label{eq:g2}\\
& &\!\!\!\!\!\!\!\!\!\!\!\!\!\!\!\!\!\!\!\!\!\!\!\!
\left[\frac{m_V}{2p}+\Gamma_i(p,E)
- \frac{m_V}{2(p-k)}
+\Gamma_o(p-k,E-\frac{\lambda^2}{2k})
\right] \nonumber\\
\Gamma_o(p^\prime,E^\prime) &=& \frac{g^2}{2p^\prime} 
\int_0^{p^\prime} \!\!\! 
\frac{dk}{2k}
G(p^\prime-k,E^\prime-\frac{\lambda^2}{2k}) \times \label{eq:g3}\\
& &\!\!\!\!\!\!\!\!\!\!\!\!\!\!\!\!\!\!\!\!\!\!\!\!
\left[\frac{m_V}{2(p^\prime-k)}
+\Gamma_i(p^\prime-k,E^\prime-\frac{\lambda^2}{2k})
- \frac{m_V}{2p^\prime}+
\Gamma_o(p^\prime,E^\prime)
\right] \nonumber   ,
\end{eqnarray}
which have been illustrated in Fig. \ref{fig3}
\begin{figure}
\unitlength1.cm
\begin{picture}(15,6.5)(1.5,-8)
\includegraphics{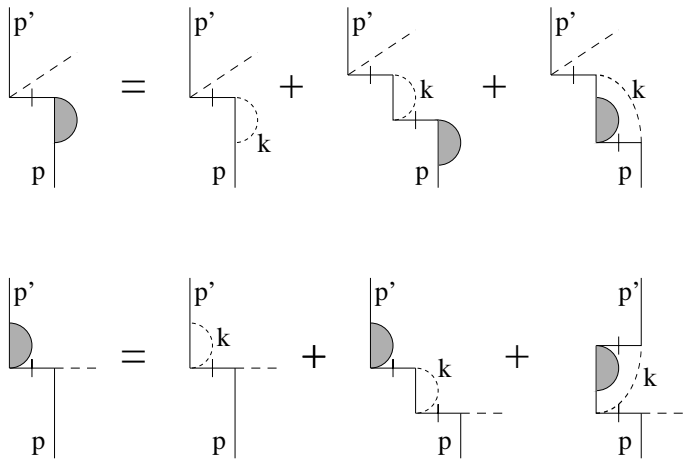}
\end{picture}
\caption{Recursive relation between loop corrections 
for the dressed vertex.}
\label{fig3}
\end{figure}

Given the Green's function, one can thus calculate $\Gamma $
nonperturbatively, by solving Eqs. (\ref{eq:g1}-\ref{eq:g3})
self-consistently (e.g. by iteration).

\begin{figure}
\unitlength1.cm
\begin{picture}(15,3)(0,-4.3)
\includegraphics{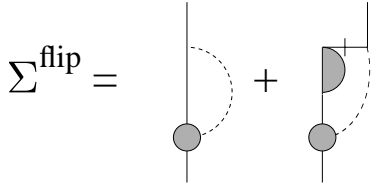}
\end{picture}
\caption{Self-energy generated by two consecutive (dressed) helicity
flip interactions. For the second vertex, dressing on the ``outside''
has to be excluded in order to avoid double counting.
}
\label{fig4}
\end{figure}

\begin{figure}
\unitlength1.cm
\begin{picture}(15,4.5)(1.1,-5.5)
\includegraphics{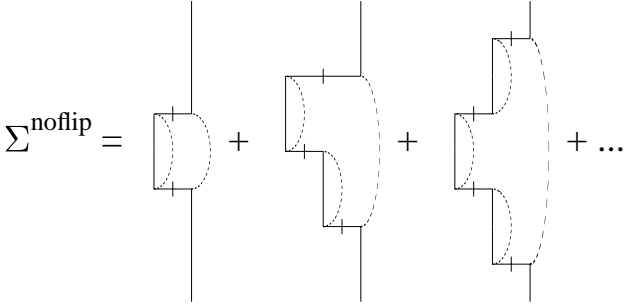}
\end{picture}
\caption{
Self energy diagrams without helicity flip.
}
\label{fig5}
\end{figure}

In order to calculate the Green's function $G(p,E)$, it is useful to
divide up all self energies $\Sigma$ into two main classes:
those which do not involve helicity flip (except in sub-diagrams) 
$\Sigma^{no\, flip}$
and those where the outermost ``gluon'' loop does involve double helicity 
flip $\Sigma^{flip}$.
\begin{equation}
\Sigma (p,E) = \Sigma^{flip}(p,E)+\Sigma^{no\, flip}(p,E) .
\end{equation}
They satisfy (Fig. \ref{fig4})
\begin{eqnarray}
\Sigma^{flip}(p,E) &=& g^2 \int_0^p \frac{dk}{k} 
\Gamma(p,p-k,E,E-\frac{\lambda^2}{2k})
\times  \\
& & G(p-k,E-\frac{\lambda^2}{2k}) \times \nonumber\\
& &\left[ \frac{m_V}{2p}-\frac{m_V}{2(p-k)} 
+ \Gamma_i(p-k,E-\frac{\lambda^2}{2k}) \right] \nonumber
\end{eqnarray}
as well as (Fig. \ref{fig5})
\begin{eqnarray}
\Sigma^{no\, flip} &=& \frac{g^2}{2} \int_0^p \frac{dk}{(p-k)k} 
\frac{\Sigma^{di}(p-k,E-\frac{\lambda^2}{2k})}
{1-\Sigma^{di}(p-k,E-\frac{\lambda^2}{2k})}
\label{eq:noflip}
\end{eqnarray} 
where the ``double instantaneous'' self energy $\Sigma^{di}$
is given by 
\begin{equation}
\Sigma^{di}(p,E) = \frac{g^2}{2p}\int_0^p \frac{dk}{k}
G(p-k,E-\frac{\lambda^2}{2k}) .
\label{eq:sdi}
\end{equation}
Finally, one can calculate the Green's function from the self-energies
using
\begin{equation}
G(p,E) = \frac{1}{E - \frac{m_{kin}^2}{2p} - \Sigma(p,E)} .
\label{eq:G}
\end{equation}
The above set of equations allows one to determine both self-energy
and vertex functions in a self-consistent manner. 
The solution to these integral
equations is completely non-perturbative and should thus contain the
physics of D$\chi$SB.

\section{An Illustrative Example: $\lambda \rightarrow \infty$}
In general the coupled integral equations relating vertex and self-energy
need to be solved numerically. However, in this section we will illustrate the
physics of these equations by considering the limiting case where the
``gluon'' mass is very large and thus all loop diagrams become energy
independent --- the Gross-Neveu limit.

For example, the Green's function becomes that of a free particle
\be
G(p,E) = \frac{1}{E-\frac{M^2_{phys}}{2p}+i\varepsilon}.
\label{eq:Gfree}
\ee
Below we will show this result explicitly and we will also determine
$M_{phys}$ self-consistently. For $\lambda \rightarrow \infty$, not
only self energies, but also the ``vertex-corrections'' $\Gamma$ become
energy independent and one can thus parameterize them in the form
\be
\Gamma_i(p,E)=-\Gamma_o(p,E) = \frac{\Delta M_V}{2p} .
\label{eq:defMV}
\ee
From Eqs. (\ref{eq:g2}) one finds for example
\bea
\Gamma_i(p,E) &=&\frac{g^2}{2p} \int_0^1 \frac{dx}{1-x}
\frac{M_V}{\frac{M_{phys}^2}{1-x}+ \frac{\lambda^2}{x} - 2pE}
\nonumber\\
&\stackrel{\lambda^2 \rightarrow \infty}{\longrightarrow}&
\frac{M_V}{2p} \frac{g^2}{\lambda^2} \ln \frac{\lambda^2}{M_{phys}^2}  
\label{eq:Gammalimit}
\eea 
where $M_V \equiv m_V+\Delta M_V$, and where we take the limit
$\lambda^2 \rightarrow \infty$ in such a way that
$\frac{g^2}{ \lambda^2} \ln (\lambda^2/M_{phys}^2)$ remains fixed in order to obtain
a nontrivial limit. Thus, combining Eq. (\ref{eq:defMV}) with Eq.
(\ref{eq:Gammalimit}), one finally obtains
\be
M_V = m_V + \frac{M_Vg^2}{\lambda^2} \ln \frac{\lambda^2}{M_{phys}^2} .
\label{eq:MV},
\ee
i.e.
\be
M_V = \frac{m_V}{1 -\frac{g^2}{\lambda^2} \ln \frac{\lambda^2}{M_{phys}^2} }.
\label{eq:geo}
\ee
This result can now be used to evaluate the self-energy and hence the
physical mass $M_{phys}$. 
In order to avoid having to deal with any zero mode degrees of freedom,
we implicitly use a Pauli-Villars (PV) regulator, where the kinetic mass 
counter-term induced by zero-modes is known to vanish (see for example
Refs. \cite{mb:rot,hiller}).

First of all one finds that the ``double-instantaneous'' self-energy
$\Sigma^{di}$ (\ref{eq:sdi})
vanishes for $\lambda^2 \rightarrow \infty$, since it does not
contain the $\ln \lambda^2$.
\be
\Sigma^{di} \stackrel{\lambda^2 \rightarrow \infty}{\longrightarrow}
\frac{g^2}{2p \lambda^2} \propto \frac{1}{\ln \frac{\lambda^2}{M^2_{phys}}}
\stackrel{\lambda^2 \rightarrow \infty}{\longrightarrow } 0
\ee
and thus $\Sigma^{no\, flip}$ (\ref{eq:noflip}) can be neglected.

For $\Sigma^{flip}$, which is the only divergent self-energy in this model,
a PV regulator is used at intermediate steps. 
\footnote{An alternative, which we will not pursue in this paper,
would be to include a separate kinetic mass counter-term.}
After a few trivial algebraic steps \cite{mb:rot,hiller}, one eventually finds
\bea
\Sigma^{flip}&=& \frac{g^2}{2p} \int_0^1 \frac{dx}{x}
\frac{ \left(M_V-\frac{M_V}{1-x}\right)
\left(m_V-\frac{M_V}{1-x}\right) }
{2pE - \frac{M_{phys}^2}{1-x} - \frac{\lambda^2}{x}}
\nonumber\\
&\stackrel{PV}{\longrightarrow}& -
\frac{g^2}{2p} \int_0^1 \frac{dx}{x}
\frac{\left[m_VM_V\frac{x}{1-x} + \frac{M_V^2}{1-x}
+2pE\right]}{2pE - \frac{M_{phys}^2}{1-x} - \frac{\lambda^2}{x}}
\nonumber\\
&\stackrel{\lambda^2 \rightarrow \infty}{\longrightarrow}&
M_V \left( m_V+M_V\right) 
\frac{g^2}{2p\lambda^2} \ln \frac{\lambda^2}{M_{phys}^2}, 
\eea
which is energy independent and thus confirms the original ansatz 
for $G(p,E)$ Eq. (\ref{eq:Gfree}). Using Eqs. (\ref{eq:G}) and (\ref{eq:Gfree})
one thus finds
\be
M_{phys}^2 = m_V^2 
+ M_V \left( m_V+M_V\right) 
\frac{g^2}{\lambda^2} \ln \frac{\lambda^2}{M_{phys}^2} 
\label{eq:Mphys}.
\ee
Note that we made use here of the fact that the kinetic mass $m_{kin}$
in the Hamiltonian 
and the bare vertex mass $m_V$ are identical when a PV regulator is employed
\cite{mb:rot,hiller}.

In order to disentangle the relation between bare vertex mass $m_v$,
dynamical vertex mass $M_V$ and the physical mass $M_{phys}$, we
use Eq. (\ref{eq:MV}) to eliminate the logarithm in Eq. (\ref{eq:Mphys}),
yielding
\be
M_{phys}^2 = m_V^2 + M_V \left(m_V+M_V\right) \frac{\left(M_V-m_V\right)
}{M_V} = M_V^2,
\ee
i.e. in this simple example, the consistent solution of the coupled system of equations for
the vertex function and the self-energy yields a physical mass which
is identical to the dynamical vertex mass.
Note that this equality of $M_{phys}$ 
(defined through the lowest eigenvalue of 
the invariant mass operator) and $M_V$ (defined through the helicity 
flip amplitude) is a nontrivial result. In more general models, one
would still expect that there are some connections between $M_{phys}$
(or a constituent mass) and $M_{phys}$ but they do not necessarily
have to be numerically equal.

Furthermore, if we use this identity between $M_{phys}$ and $M_V$ in
Eq. (\ref{eq:MV}) [or in Eq. (\ref{eq:Mphys})], one finds that
\be
M_V = m_V + M_V\frac{g^2}{\lambda^2} \ln \frac{\lambda^2}{M_V^2} 
\ee
which is not only identical to the covariant ``gap equation'' for this
model, but which also admits a nonzero solution for $M_V$ in the chiral 
limit ($m_V \rightarrow 0$)
\be
M_{phys}=M_V = \lambda e^{-\frac{\lambda^2}{2g^2}} ,
\label{eq:MVdyn}
\ee
i.e. a nonzero vertex mass has been generated dynamically.
Obviously, Eq. (\ref{eq:MVdyn}) is non-perturbative in the coupling constant,
but this should not be too surprising since, even though we used
diagrammative techniques, what we have done is in fact solved a
bound state problem and bound states are always non-perturbative.

From Eq. (\ref{eq:geo}) one would naively expect that the physical
vertex mass $M_V$ vanishes in the chiral limit, i.e. when the bare
vertex mass $m_V$ vanishes. However, a glance at Eq. 
(\ref{eq:MVdyn}) reveals that the denominator of Eq. (\ref{eq:geo})
also vanishes in the chiral limit. When one performs a perturbation 
expansion of the physical helicity flip amplitude in the coupling $g$,
one finds that every single diagram vanishes in the chiral limit,
since every helicity flip diagram contains at least one helicity
flip interaction which is proportional to $m_V$. However, upon
summing over an infinite number of diagrams, one obtains a divergent
geometric series which compensates for the vanishing of $m_V$.
In fact, substituting $M_{phys}$ for $M_V$ in Eq. (\ref{eq:Mphys}),
one finds that
\be
1-\frac{g^2}{\lambda^2} \ln \frac{\lambda^2}{M_{phys}^2}
= \frac{m_V}{M_{phys}}
\label{eq:den0}
\ee
which explicitly shows the vanishing of the denominator in
Eq. (\ref{eq:geo}) for $m_V \rightarrow 0$.

From the practitioner's point of view, it is very important to know 
how a dynamically generated vertex mass would in principle emerge in a 
Fock space expansion.
In order to elucidate this point, let us suppose that we know
$M_{phys}$, but not $M_V$ and let us attempt to determine $M_V$
perturbatively. For this purpose, let us
consider a helicity flip process and focus on all those diagrams which
are linear in the vertex mass, i.e. contain only one helicity flip vertex.
Since we suppose that we know the physical mass, it makes sense to
redefine the kinetic mass as the physical mass, but then one should no
longer take into account diagrams with self-mass sub-diagrams 
\footnote{I.e. sub-diagrams 
with only two external fermion lines, which are both on mass shell.}
since this would amount to double counting. Without self-mass
sub-diagrams and in the planar approximation considered here, this leaves 
only a very simple class of diagrams which are linear in the vertex mass.
In the following, we will study this class in detail and sum it up to all
orders in perturbation theory.

First of all, since we work in a dimensionally reduced model,
boson-fermion vertices are linear in the vertex mass, unless they
are instantaneous vertices, i.e. the limitation to terms linear in the vertex
mass means that all but one vertex in a given time ordered diagram 
must be instantaneous. This means that, to leading order in the vertex mass, 
a general higher order diagram
for a helicity flip vertex is obtained by attaching an instantaneous
interaction to the external vertex and then another one to the boson line
emanating from the first instantaneous vertex and so on until the last
boson line ends up in the actual helicity flip vertex.
Typical diagrams are depicted in Fig. \ref{fig1} 
\footnote{Of course, in addition to
the diagrams in Fig. \ref{fig1}, there are corrections to the helicity
flip amplitude on the other side of the external vertex, but up to
kinematical factors, these are identical to the ones depicted in
Fig. \ref{fig1}.}

Let us first look at the ``bare rainbow'' diagrams \footnote{We call these
diagrams bare rainbow diagrams, since they really have the topology of
a rainbow, as distinguished from iterated or nested rainbows.}
in the top row of Fig. \ref{fig1}. 

\begin{figure}
\unitlength1.cm
\begin{picture}(15,6.5)(1,-8.5)
\includegraphics{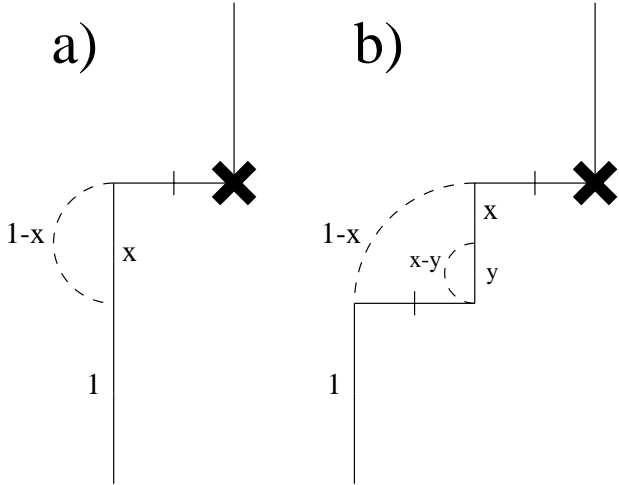}
\end{picture}
\caption{Lowest order bare rainbow diagrams.}
\label{fig:bare}
\end{figure}

\begin{figure}
\unitlength1.cm
\begin{picture}(15,3)(1.3,-4)
\includegraphics{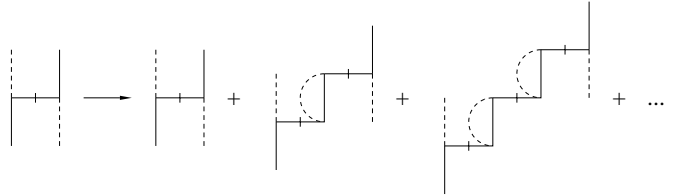}
\end{picture}
\caption{Chain of instantaneous interactions leading to a geometric
series.
}
\label{fig:chain}
\end{figure}

We will restrict ourselves to vertex corrections in the incoming 
state. Furthermore, in order to simplify the notation, we absorb a trivial 
kinematic factor
into the definition of the helicity-flip amplitude by introducing
\be
T_{flip} \equiv 2p \Gamma_{i} .
\ee
The first diagram in Fig. \ref{fig1} is just the bare vertex.
The second diagram yields
(Fig.\ref{fig:bare}a)
\begin{equation}
T_{flip}^{2a}=
m_Vg^2 I_1,
\label{eq:t2}
\end{equation}
where
\begin{eqnarray}
&I_1 &= \int_0^1 \frac{dx}{(1-x)} \frac{ \frac{1}{x}-1}
{2pE-\frac{M^2_{phys}}{x}-\frac{\lambda^2}{1-x}}
\nonumber\\ & &
\stackrel{\lambda \rightarrow \infty}{\longrightarrow}
\frac{1}{\lambda^2} \ln \frac{\lambda^2}{M^2_{phys}}.
\end{eqnarray}
The fourth order bare rainbow is a little more complicated, yielding
(up to the same overall factors as Eq. (\ref{eq:t2})
(Fig. \ref{fig:bare}b)
\begin{eqnarray}
& &T_{flip}^{2b}=g^4m_V \int_0^1 \frac{dx/({1-x})}
{2pE-\frac{M^2_{phys}}{x}-\frac{\lambda^2}{1-x}}  \times
\label{eq:t4}
\\
& & \quad \quad \quad \quad\quad \quad \quad \quad
\quad \quad \quad 
\int_0^x \frac{dy/({x-y})\left(\frac{1}{y}-\frac{1}{x}\right)}
{2pE-\frac{M^2_{phys}}{y}-\frac{\lambda^2}{x-y}-\frac{\lambda^2}{1-x}}
.\nonumber
\end{eqnarray}
The $x$ integral  in Eq. (\ref{eq:t4}) is
dominated by small values of $x$ as $\lambda \rightarrow \infty$.
This allows us to simplify the $y$ integral ($z=y/x$)
\begin{eqnarray}
\int_0^x \frac{dy}{x-y} \frac{\left(\frac{1}{y}-\frac{1}{x}\right)}
{2pE-\frac{M^2_{phys}}{y}-\frac{\lambda^2}{x-y}-\frac{\lambda^2}{1-x}}
\!\!\!\!\!\!\!\!\!
\!\!\!\!\!\!\!\!\!
\!\!\!\!\!\!\!\!\!
\!\!\!\!\!\!\!\!\!
\!\!\!\!\!\!\!\!\!
\!\!\!\!\!\!\!\!\!
\!\!\!\!\!\!\!\!\!
& & 
\\ & &
=\int_0^1 \frac{dz}{1-z} \frac{\left(\frac{1}{z}-1\right)}
{x\left(2pE-\frac{\lambda^2}{1-x}\right)-\frac{M^2_{phys}}{z}
-\frac{\lambda^2}{1-z}}
\nonumber\\
& &
\stackrel{x \rightarrow 0}\longrightarrow
-\int_0^1 \frac{dz}{1-z} \frac{\left(\frac{1}{z}-1\right)}
{\frac{M^2_{phys}}{z}+\frac{\lambda^2}{1-z}}
\stackrel{\lambda \rightarrow \infty}\longrightarrow -I_1 ,
\nonumber
\end{eqnarray}
i.e.
\begin{equation}
T_{flip}^{2b}= m_V \left(g^2I_1\right)^2.
\end{equation}
It turns out that the bare rainbow diagrams form a geometric series,
yielding (together with the bare vertex)
\begin{equation}
T_{flip}^{bare \ rainbow}=m_V\frac{1}{1-g^2 I_1} =
m_V\frac{1}{1-\frac{g^2}{ \lambda^2}\ln 
\frac{\lambda^2}{M^2_{phys}}}.
\label{eq:tbare}
\end{equation}
Besides the bare rainbow diagrams one also needs to consider the
nested rainbow diagrams (bottom row of Fig. \ref{fig1}), where
the the direction of the chain of instantaneous interactions does
not always alternate along the fermion line. These can be obtained
from the bare rainbow diagrams by replacing each instantaneous
interaction by a chain of instantaneous interactions (Fig. \ref{fig:chain})
leading again to a geometric series as $\lambda \rightarrow \infty$
\footnote{Only for $\lambda \rightarrow \infty$ can one neglect the energy
dependence of these bubbles.}, which can be incorporated into above
result by making the replacement
\begin{equation}
g^2 \longrightarrow \frac{g^2}{1- g^2I_2},
\label{eq:geo2}
\end{equation}
where
\begin{equation}
I_2 = \int_0^1 \frac{dx}{(1-x)} \frac{1}
{p^2-\frac{M^2}{x}-\frac{\lambda^2}{1-x}}
\stackrel{\lambda \rightarrow \infty}{\longrightarrow}
\frac{1}{\lambda^2}
\end{equation}
in each instantaneous interaction. Together with Eq. (\ref{eq:tbare}),
this yields for the sum of all planar helicity flip diagrams in leading
order in the vertex mass
\begin{equation}
T_{flip}^{rainbow}=\frac{m_V}{1-g^2 \frac{I_1}{1- g^2 I_2}} .
\label{eq:tfull}
\end{equation}
However, since 
\be
I_2/I_1 \sim 1/\ln \lambda^2 
\stackrel{\lambda \rightarrow \infty}{\longrightarrow}0
\label{eq:ratio}
\ee
, one can neglect
$I_2$ in Eq. (\ref{eq:tfull}), yielding
\begin{equation}
T_{flip}^{rainbow} \
\stackrel{\lambda \rightarrow \infty}{\longrightarrow}
\frac{m_V}{1-\frac{g^2}{ \lambda^2}\ln \frac{\lambda^2}{M^2}},
\label{eq:tflip}
\end{equation}
i.e. just the bare result above (\ref{eq:tbare}).

From our calculation above (\ref{eq:den0}), we know already that the perturbation
series diverges in the chiral limit. Eq. (\ref{eq:ratio}) tells us that
the divergence arises from summing over all ``bare rainbow'' diagrams
(Fig. \ref{fig:bare}).

The crucial point is that the denominator of Eq. (\ref{eq:tflip}) becomes 
small in the chiral limit, as one can read off from Eq. 
(\ref{eq:den0}). In other words, even though each individual diagram
in Fig. (\ref{fig1}) vanishes in the chiral limit $\propto m$, 
summing over an infinite number of diagrams yields 
$
T_{flip} \propto M,
$
which remains finite as $m\rightarrow 0$.

Several interesting observations can be made from this example:
\begin{itemize}
\item While zero-modes contribute significantly to the kinetic mass
of fermions, there is no such zero-mode contribution to the vertex mass.
However, the vertex mass gets renormalized by (infinitesimally)
small $x$ contributions since one
has to sum over an infinite number of Fock space components 
in order to obtain a finite helicity flip amplitude in the chiral limit.
\item In realistic non-perturbative calculations of hadron spectra, where one
cannot include an infinite number of Fock space components, it will at some
level be necessary to absorb the higher order corrections into an effective
vertex mass $M$.
\item The leading diagrams (top row in Fig. \ref{fig1}) have a relatively
simple structure: higher order diagrams can be successively built by
replacing the bare helicity flip vertex inside a given diagram with
the second order dressed helicity flip vertex. This observation may be
useful for a renormalization group study of the helicity flip interactions,
since a large amplitude is obtained only through an infinite chain of steps
from finite $x$ down to vanishingly small values of $x$.
Qualitatively, this mechanism resembles the chiral symmetry breaking 
mechanism suggested in Ref. \cite{lenny}. 
\end{itemize}

\section{DMVG in a Covariant Framework}
Even though we have formulated the mechanism for DVMG entirely in the
Hamiltonian LF framework, it is very instructive to rephrase
the above results using covariant language. It should be emphasized that 
all results derived in this section are of course
also contained in the LF-Hamiltonian result.

For the above model, the nonperturbatively obtained covariant
self-energy (obtained for example by solving Dyson-Schwinger 
equations) can be expressed in the form
\be
\Sigma = A(p^2) \not \! \! \,p + B(p^2) ,
\label{eq:covsigma}
\ee
yielding for the dressed fermion propagator
\bea
S &\equiv& \frac{1}{\not \! \! \,p - m - \Sigma} =
\frac{1}{\not \! \! \,p \left(1-A(p^2)\right)- m - B(p^2)}
\nonumber\\
&=& \frac{\not \! \! \,p \left(1-A(p^2)\right)+ m + B(p^2)}
{p^2 \left(1-A(p^2)\right)^2 - \left(m + B(p^2)\right)^2},
\label{eq:Propcov}
\eea
where $\not \! \! \,p \equiv \gamma^+p^-+\gamma^-p^+$ and
$p^2=2p^+p^-$.

Dynamical chiral symmetry breaking manifests itself as $B \neq 0$
even for vanishing current quark mass $m\rightarrow 0$.
In order to compare the covariant approach with the Hamiltonian LF
framework, we ask: 
what is the helicity flip amplitude which is in this case generated by
self-energy diagrams connected to the external vertex by 
LF-``instantaneous'' $\gamma^+/p^+$ propagators?
\begin{figure}
\unitlength1.cm
\begin{picture}(15,6)(1,-7)
\includegraphics{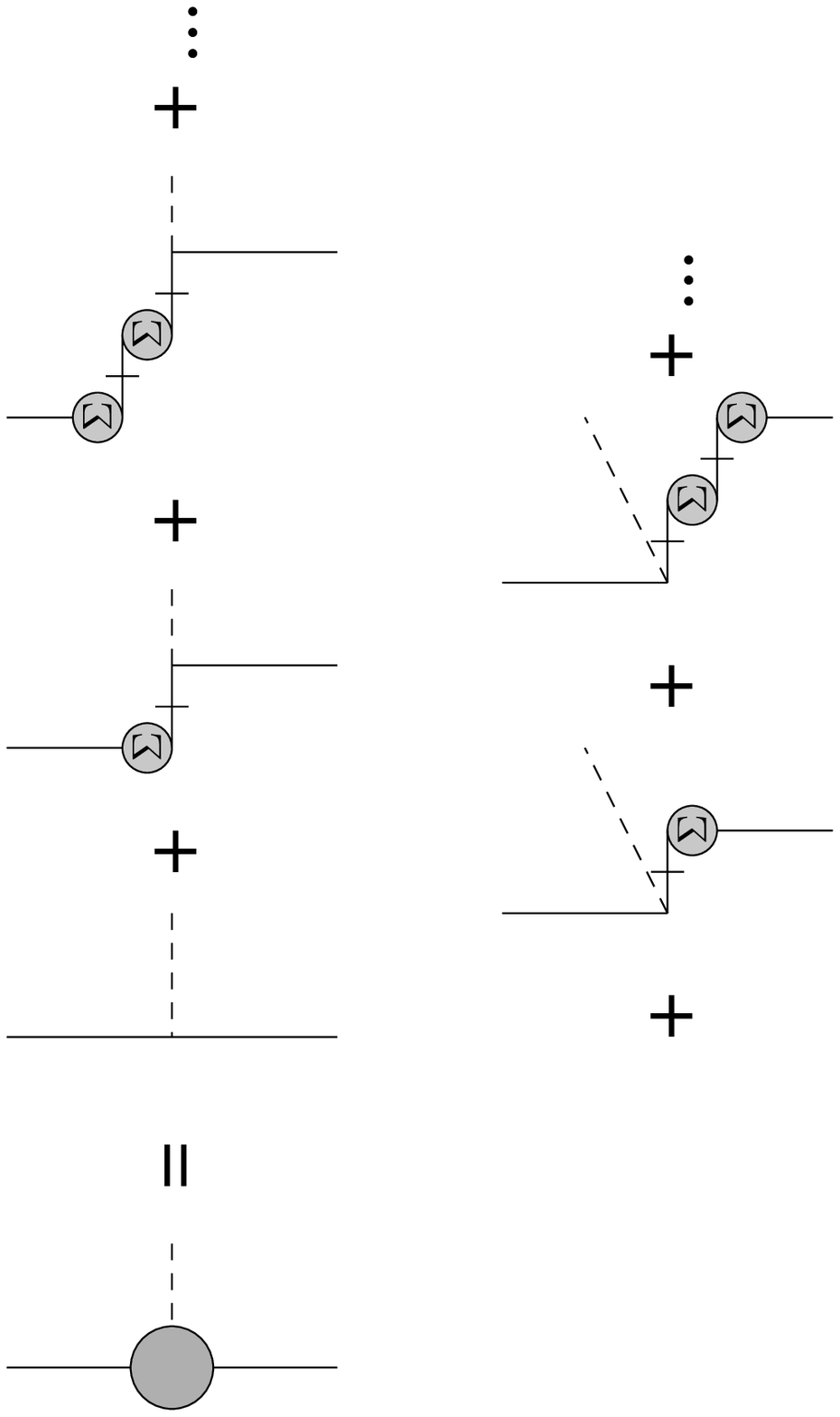}
\end{picture}
\caption{Relation between the dressed LF vertex (large blob) and
the covariant self-energy $\Sigma$ (smaller blobs).
}
\label{fig:vcov}
\end{figure}
The resulting chain of self energy diagrams (Fig. \ref{fig:vcov}) 
is a geometric
series and one thus finds for example for the sum of all corrections
attached to the incoming line
\bea
\Gamma_i(p^+,p^-) &=& \bar{u}^\prime \gamma_x \frac{\gamma^+}{2p^+} \Sigma u
+ \bar{u}^\prime \gamma_x \frac{\gamma^+}{2p^+} \Sigma \frac{\gamma^+}{2p^+} 
\Sigma u \nonumber\\ & &+ \bar{u}^\prime \gamma_x\frac{\gamma^+}{2p^+} \Sigma \frac{\gamma^+}{2p^+} \Sigma \frac{\gamma^+}{2p^+} 
\Sigma u + ...
\nonumber\\ &=& \bar{u}^\prime \gamma_x \frac{\gamma^+}{2p^+} \Sigma u
\frac{1}{1-A(p^2)}\nonumber\\ &=& 
\frac{1}{2p^+} \frac{m A(p^2) + B(p^2)}{1-A(p^2)} .
\label{eq:Gcov1}
\eea
Note that the spinor basis in which the LF Hamiltonian is constructed
contains the current quark mass. Therefore, when taking spinor
matric elements of the covariant amplitude we must also use
spinors $u$ and $\bar{u}^\prime$
which also contain the current quark mass in order to be able to relate
the covariant calculation to the LF-Hamiltonian calculation. 

The total dressed vertex (including the bare piece as well as
renormalizations on both the incoming and the outgoing line) thus reads
\bea
\Gamma &=& \frac{1}{2p^+}\left( m +\frac{m A(p^2) + B(p^2)}{1-A(p^2)} \right)
\nonumber\\
& & 
- \frac{1}{2{p^+}^\prime}\left(m +\frac{m A({p^2}^\prime) + B({p^2}^\prime)}
{1-A({p^2}^\prime)}\right)
\nonumber\\
&=&\frac{1}{2p^+}\left( \frac{m + B(p^2)}{1-A(p^2)} \right)
 - \frac{1}{2{p^+}^\prime}\left(\frac{m + B({p^2}^\prime)}
{1-A({p^2}^\prime)}\right).
\label{eq:Gcov2}
\eea
In order to understand the physics of Eq. (\ref{eq:Gcov2}), let us first 
consider the case where both the incoming and the outgoing line
correspond to on-shell states, i.e. $p^2={p^2}^\prime = M_{phys}^2$.
Analyzing the pole structure of the fermion propagator Eq. (\ref{eq:Propcov}) 
yields
\be
M_{phys}\left(1-A(M_{phys}^2)\right) = B(M_{phys}^2)
\ee
and thus
\be
\Gamma = \frac{M_{phys}}{2p^+}-\frac{M_{phys}}{2{p^+}^\prime}.
\quad \quad \quad (p^2={p^2}^\prime = M_{phys}^2) .
\label{eq:Gammaphys}
\ee
The physics of Eq. (\ref{eq:Gammaphys}) is that the physical
helicity flip amplitude in this model for on-shell fermions
is identical to the one for free (non-dressed) fermions
with a mass equal to the physical mass of the fermions.
\footnote{Note that in similar situations in the ET approach,
it is convenient to perform a Bogoliubov transformation
\cite{eric}, which yields large helicity flip
amplitudes due to self-energies in a more direct way than in
the LF approach.}

Of course, this result changes when the fermions are far from
their mass shell. In asymptotically free models, where $A(p^2), 
B(p^2) \rightarrow 0$ 
as $p^2 \rightarrow - \infty$ one recovers the current quarks
\be
\Gamma \longrightarrow \frac{m}{2p^+}-\frac{m}{2{p^+}^\prime}.
\quad \quad \quad (p^2={p^2}^\prime \rightarrow - \infty)
\label{eq:Gammaeucl} .
\ee
Note that the dressed vertex in this model depends only on the
$p^2$ of the initial and final state, but it does {\it not} depend
on the momentum transfer. This feature is of course a peculiarity
of the model which arises because we make a planar approximation
and because the ``gluons'' in the model have no direct self-interaction.
This features prohibits ``gluon'' lines that run across the ``gluon''
vertex and hence the only ``vertex'' corrections are, from a covariant
point of view, corrections to the propagators. Therefore, it is quite
natural in this model that the ``vertex corrections'' depend only on 
the invariant mass of the external fermion lines.

Another interesting observation one can make is about the connection 
between the effective mass and condensates. Obviously, the numerator 
appearing in the
dressed (helicity flip) vertex (\ref{eq:Gcov2}) is identical to the
numerator of the Dirac-trace of the dressed propagator
(\ref{eq:Propcov}). On the other hand, the trace of the propagator
appears in the expression for the Fourier transform of non-local
quark condensates, which suggests that
there might be a deep connection between the effective vertex mass
and nonlocal quark condensates.\footnote{For a recent study of
non-local condensates, see for example Ref. \cite{cmu}.} 
To illustrate this point, let us consider a
situation where $A(p^2) \ll 1$ and where $p^2 \gg \left(m+B(p^2)\right)^2$.
In this limit one obviously finds that the running vertex mass
$M_V(p^2) \equiv \frac{m+B}{1-A}$ satisfies
\be
M_V(p^2) = p^2 \mbox{tr} \left(S(p)\right) .
\ee
Intuitively, one would also expect a relation between the running
vertex mass and the local condensate. Even though we were able to
establish such a relation from the integral equations determining
the dressed vertex in certain limiting cases, these results did
depend on model details and will thus not be displayed here.

\section{Summary and Outlook}
We have studied the non-perturbative enhancement of the fermion helicity
flip amplitude in a simple model formulated in the LF framework.
The dimensionally reduced model, 
consisting of fermions coupling to the transverse
component of a vector field, treated in planar approximation
(``large $N_C$''), is of course quite different from real QCD,
and one cannot consider it an approximation to QCD or a
phenomenological model for QCD. What the model has in common with QCD
is that it is based on a chirally invariant Lagrangean and
that the chiral symmetry is dynamically broken, leading for
example to mass generation for the fermions in the limit of
vanishing current quark masses. It is because of this similarity
that we believe that studying its features in the LF framework
may shed some light on how dynamical chiral symmetry breaking
might perhaps arise in LFQCD.

In this model, we were able to show that even though every perturbative
diagram for helicity flip amplitudes is proportional to at least one
power of the vertex mass (=current mass), and thus vanishes in the chiral
limit, the sum over all diagrams yields a result which remains finite
for vanishing vertex mass. In a solvable model, which resembles the
Gross-Neveu model, we demonstrate the delicate interplay between
kinetic mass, vertex mass and physical mass, which leads to the
crucial divergence in the helicity amplitude with the vertex mass factored
out. This is necessary to counterbalance the vanishing of the vertex mass,
and we explicitly demonstrate that the physical helicity flip amplitude
stays finite in the chiral limit.

The calculation was done starting from the canonical LF-Hamiltonian
plus a kinetic mass counter-term for the fermions. Even though most
of the analysis used the language of diagrams, it should be emphasized
that we derived non-perturbative relations between Green's functions.
The results thus obtained are therefore equivalent to results that
one could also obtain by nonperturbatively solving eigenvalue equations
resulting in an infinite number of coupled Fock space equations. 

Upon analyzing the physical helicity flip amplitude,
one finds that a large amplitude in the chiral limit emerges only after 
an infinite number 
of Fock space components have been included. One can draw a number of 
lessons from this result. First, it seems that a calculation based on a
canonical Hamiltonian (plus $m_{kin}^2$ counter-term) is in principle
sufficient to describe a situation where chiral symmetry is dynamically
broken. Secondly, from a fundamental point of view, while zero-mode 
induced corrections seem crucial for kinetic mass generation and thus
also for physical mass generation (in the sense of eigenvalue of the
invariant mass operator), they are not important for dynamical vertex mass
generation (DVMG), i.e. for large physical helicity flip amplitudes and large
hyperfine splittings). Instead DVMG comes from high Fock components and thus
also the small $x$ region. From the practitioner's point of view our
results imply that a calculation based on a canonical Hamiltonian
(plus $m_{kin}^2$ counter-term) must include a very large number of
Fock components in the small fermion mass limit in order to generate 
the physics of dynamical chiral symmetry breaking.
Since a sufficiently large number of Fock components may be very difficult
to handle adequately in numerical calculations (both using DLCQ or
basis functions methods), it may thus be advantageous to improve the
canonical (plus $\Delta m_{kin}^2$) Hamiltonian by adding non-canonical
terms. In principle, these terms can be obtained by integrating out 
(or parameterizing) higher Fock components self-consistently. 

Given the abovementioned caveats, what can we learn from these results
about dynamical chiral symmetry breaking in LFQCD?
Of course, this part is only speculation, but there are some very
suggestive conjectures that one can make concerning this issue:
Several models and approximations to LFQCD, such as
dimensionally reduced models \cite{dalley} or the transverse lattice
\cite{bardeen}, give rise to effectively 1+1 dimensional theories
\footnote{It should be emphasized that the transverse lattice is still
a 3+1 dimensional formulation of QCD. However, since the transverse directions
are discretized in this approach, it can formally be regarded as a large number
of coupled 1+1 dimensional theories!}. These models or approximations
contain both interactions which resemble interactions in $QCD_{1+1}$ 
as well as couplings to the transverse gauge degrees of freedom that very 
much resemble the couplings of the model studied in this paper.
As has been pointed out in Ref. \cite{dalley}, the end-point behavior
of wave functions for higher Fock components is dominated by the couplings
of the fermions to the transverse gauge degrees of freedom and {\it not}
by the longitudinal gauge interaction. Since the
non-perturbative enhancement of helicity flip enhancement described in this
paper originated in the small $k^+$ and high Fock component region, one would
thus expect that the same or a similar mechanism also works in
collinear QCD models as well as on the transverse lattice.
If this is indeed the case then it is also clear that numerical techniques
based on a Fock space expansion (both DLCQ as well as basis function methods)
are doomed to fail in the chiral limit, unless effective interactions
are added which mimic the effects from the small $k^+$ and high Fock component
region,  which typically has to be omitted for practical reasons.
The most simple operator that one might think of based on the results
of this paper is a ``running vertex mass''. Integrating out 
small $k^+$ and high Fock components should lead to effective quarks with the
following properties: at low momentum transfers and with initial and final
states not too far off shell, the quarks in collinear QCD or $\perp$ lattice
approaches should acquire a large (constituent mass scale!) effective
vertex mass. However, because of asymptotic freedom, one should recover the
current quarks when the states are highly off shell.
Such features can be incorporated into an effective LF Hamiltonian
with a ``running'' vertex mass. A detailed discussion of this running, 
which will depend on the precise nature of the model and of the 
cutoff employed, would go beyond the intended scope of this paper.

\acknowledgements
I would like to thank the participants of the ``International Workshop
on Light-Cone QCD and Nonperturbative Hadronic Physics'' in
Lutsen, MN (August 97) for many inspiring discussions. I would also
like to thank Bob Klindworth for reading and criticizing the manuscript.
This work was supported by the D.O.E. (grant no. DE-FG03-96ER40965)
and in part by TJNAF.

\end{document}